\documentclass[showpacs, onecolumn,12pt]{revtex4-1}

\usepackage{amsbsy}
\usepackage{amsfonts}
\usepackage{amssymb}
\usepackage{amsmath}
\usepackage{pdfpages}
\usepackage{graphicx}

\begin{document}
\title{Do entangled photons
induce \\ two-photon two-atom transitions \\ more efficiently than other states of light  ?}

 \author{Zhan Zheng}
\affiliation{Laboratoire Kastler Brossel,  Universit\'e Pierre et Marie Curie-Paris 6,\\
ENS, CNRS; 4 place Jussieu, 75252 Paris, France and\\
State Key Laboratory for Precision Spectroscopy,
East China Normal University, Shanghai 200062, China}

\author{ Pablo L. Saldanha}
\affiliation{Departamento de F\'isica, Universidade Federal de Pernambuco, 50670-901, Recife, PE, Brazil and\\ Departamento de F\'isica, Universidade
Federal de Minas Gerais, Caixa Postal 702, 30161-970, Belo Horizonte,
MG, Brazil}

\author{Jos\'e R. Rios Leite}
\affiliation{Departamento de F\'isica, Universidade Federal de Pernambuco, 50670-901, Cidade Universit\'{a}ria, Recife, PE, Brazil }

\author{Claude Fabre}
\affiliation{Laboratoire Kastler Brossel,  Universit\'e Pierre et Marie Curie-Paris 6,\\
ENS, CNRS; 4 place Jussieu, 75252 Paris, France}

\begin{abstract}

We study in this paper the efficiency of different two-photon states of light to induce the simultaneous excitation of two atoms of different kinds when the sum of the energies of the two photons matches the sum of the energies of the two atomic transitions, while no photons are resonant with each individual transition. We find that entangled two-photon states produced by an atomic cascade are indeed capable of enhancing by a large factor the simultaneous excitation probability as compared to uncorrelated photons, as predicted some years ago by Muthukrishnan et al, but that several non-entangled, separable, correlated states, produced either by an atomic cascade or parametric down conversion, or even appropriate combinations of coherent states, have comparable efficiencies. We show that the key ingredient for the increase of simultaneous excitation probability is the presence of strong frequency anti-correlation and not time correlation nor time-frequency entanglement.

\end{abstract}

\today

\maketitle

\section{Introduction}

Quantum entanglement, and its inherent non local properties, are among the most fascinating and challenging features of the quantum world.
In addition, entanglement plays a central role in quantum information \cite{Plenio07,Gerardo07,Reid09,Ryszard09,Pan12}.
Since its first description in the decade of 1930 (\cite{Schrodinger1935}), and in spite of the decisive contribution of J. Bell\cite{bell1964} and the subsequent experimental studies\cite{aspect1982},
entanglement stays even now as a rather mysterious and puzzling property of bipartite quantum objects.
In particular distinguishing between effects related to genuine entanglement and those related to the quantum correlations
measured on a single quantum observable is a difficult task\cite{TrepsFabre}, as can be seen for example by the great number of papers about
quantum discord \cite{Harold, Kavan}. Some time ago,  a paper was published \cite{scully04} which discusses how entangled
states would be able to induce transitions in quantum systems that factorized states could not excite.
The physical problem studied in that paper is therefore a good test bench to examine in detail in a simple situation the role of entanglement
and of correlations not related to entanglement. This is the purpose of the present paper.

The problem under consideration is the probability of two-photon  two-atom (2P2A) excitation, in the situation where
the two atoms are of different
species and have different transition frequencies and the light to which the atoms are submitted is in general
non-resonant for each one, but resonant for the system of two atoms.
Two photon absorption by single atoms or molecules have been studied since 1931  \cite{TP-G-Mayer} and remains a current
subject of theoretical and experimental research \cite{Mollow, Bjorkholm, Fei,Lloyd, Kastella}.
When the atoms have more than one intermediate state many important features, including cross section cancellation and enhancement, are observed  \cite{Bjorkholm}. These features have recently been shown to be applicable in characterizing the quantum states of the absorbed two photon  \cite{Kastella}.

It has also been known for a long time,  that the two photons resonant excitation of two different atoms is indeed possible
when the two atoms are interacting  \cite{Jose}. A nearly monochromatic light beam will have a resonant  two photon absorption
peak when tuned across the average frequency of the two atoms. Different experiments have since then confirmed this theoretical prediction \cite{white,Sandoghdar,bagnato}.  In addition to a direct potential interaction between the atoms, like the dipole-dipole, cooperative 2P2A has also been predicted for pairs of atoms inside an optical cavity \cite{agarwal92}. In this case the physical interaction is mediated by the radiation background surrounding the atoms.
Ref \cite{scully04} addresses the case of two photon absorption
in absence of interaction between the two atoms,
with the excitation made using some particular entangled state of light. The conclusion of the authors of \cite{scully04} is
that \textit{in some situations entanglement can replace a real physical interaction}, which is a far reaching statement and an important physical property related to entanglement. Surprisingly, this question did not attract much attention during several years.
The same subject was also considered, but in the context of spin entanglement in Electron Paramagnetic Resonance by
K. Salikhov\cite{Salikhov}. More recently, \cite{Mukamel,Dayan} the related problem of interaction with pairs of broadband spectrum photons has been discussed.

In this paper, we will determine the probability of 2P2A excitation by different multi-modal states of light. From these results we draw conclusions on the respective role of entanglement and of correlations not related to entanglement in such a process. Section II gives the general framework in which  the problem is treated. Section III derives from a second order perturbation theory the relevant transition probability. Results for various different two photon states, introduced in section IV, are given in section V and VI. Finally, in section VII we discuss different hypotheses for the physical origin of the enhancement of the 2P2A process.

\section{The model}

Let us first precise the model we are using and the notations. We consider two different two-level atoms labeled (1) and (2), having ground and excited states $| g_i \rangle$ and $| e_i \rangle$ ($i=1,2$), corresponding Bohr frequencies $\omega_i$ and spontaneous emission rates $\gamma_i$ , interacting with a quantized field. We assume that  that the transitions occur in times must shorter than the two atom lifetimes so that we can consider that the two excited states have infinite lifetimes  ($\gamma_{1,2}\simeq 0$ ) and keep for ever their excitation. For the sake of simplicity we will assume that the light source is far from the atoms, so that the only non-empty modes are plane-wave modes having a single propagation direction $Oz$ and a single polarization. In this situation one can use annihilation operators depending only on the frequency $a(\omega_{\ell})=a_{\ell}$. At the rotating wave approximation the hamiltonian of the system is then given by :
  \begin{alignat}{2}\label{quant}
    H&=H_{atom}^0+H_f^0 +V &&\notag\\
    H_{atom}^0&=\hbar \omega_1 b_1^{\dag}b_1+\hbar \omega_2b_2^{\dag}b_2, \ \ b_i=|g_i\rangle\langle e_i| &&\notag\\
    H^0_f&=\sum_{\ell} \hbar \omega a_{\ell}^{\dag}a_{\ell},
    \ \ [a_{\ell}, a^{\dag}_{\ell'}]=\delta_{\ell,\ell'} & & \notag \\
 V&=\hbar b_1^{\dag} \sum_{\ell} f_{1}(\omega_{\ell})a_{\ell}+
 \hbar  b_2^{\dag} \sum_{\ell} f_{2}(\omega_{\ell})a_{\ell}+h.c.,
  \end{alignat}
where $f_i(\omega_{\ell})=-i d_i \sqrt{\omega_{\ell}/2\hbar \varepsilon_0 S L} \, e^{i \omega_{\ell} z_i/c}=f_{i\ell} e^{i \omega_{\ell} z_i/c}$, $f_{i\ell}$ being a slowly varying function of the photon frequency, can be treated as a constant $f_i\equiv f_i(\omega_i)\approx f_i(\omega_{\ell}) $. $d_i$ is the electric dipole matrix element of atom $i$, $S$ the transverse section of the beam which is focused on the atoms, $z_i$ the position of atom $i$ and $L$ the length of the quantization box, the mode density in terms of frequencies $\omega_{\ell}$ being $2 \pi c/L$. For simplicity, we will set $z_1\simeq z_2 \simeq 0$ and hence will not consider propagation effects.

The evolution of whole system is described by a unitary operator $U$, $\rho(t)=U(|g_1g_2\rangle\langle g_1g_2|\otimes\rho_0)U^{\dag}$, where $\rho_0$ is the input light state, which can be either a pure two-photon state $|\Psi_{\mu}\rangle=\sum_{kq}c_{kq}^{\mu}|1_k,1_q\rangle$ or a mixed state in its spectral decomposition form $\rho_0 =\sum_{\mu}p_{\mu}|\Psi_{\mu}\rangle\langle\Psi_{\mu}|$. The probability of 2P2A excitation is given by :
\begin{equation}\label{P}
P(t)=\sum_{\mu }p_{\mu} \langle\Psi_{\mu}|\langle g_1g_2|U^{\dag}(t) |e_1e_2\rangle\langle e_1e_2|U(t) |g_1g_2\rangle|\Psi_{\mu}\rangle
\end{equation}
so that the transition probability is known when the field variable operator $\langle g_1g_2|U(t) |e_1e_2\rangle$  is determined and the incident light state $\rho_0$ is known. The exact expression of the evolution operator is unfortunately not easy to obtain. To simplify our discussion, we will use lowest order perturbation theory, which is a good approximation for bi-photon states which do not carry much energy.

\section{second order perturbation theory}

If we assume that the coupling between the light field and the two atoms is weak, the leading term in the evolution is $U^{(2)}=-\hbar^{-2}e^{-it(H_{atom}^0+H_f^0)/\hbar}\int_0^td\tau\int_0^{\tau}ds\tilde{V}(\tau)\tilde{V}(s)$, where $\tilde{V}$ is the coupling term in the interaction picture. One has

\begin{alignat}{2}
&\langle e_1e_2|U^{(2)}|g_1g_2\rangle=e^{-it(\omega_1+\omega_2+H_f^0/\hbar)}\sum_{mn}a_ma_n\mathcal{A}_{mn}&&\\
&\mathcal{A}_{mn}=f_{1}(\omega_m) f_{2} (\omega_n)\frac{1-e^{i(\omega_1-\omega_m)t}}{\omega_m-\omega_1}
    \frac{1-e^{i(\omega_2-\omega_n)t}}{\omega_n-\omega_2}  \label{ampl}  &&
\end{alignat}
so that the leading term of the co-excitation probability (\ref{P}) reads
\begin{equation}\label{proba}
P(t)\approx \sum_{jkmn} \mathcal{A}_{jk}^*\mathcal{A}_{mn} \hbox{Tr}(a_j^{\dag}a_k^{\dag}a_ma_n \rho_0)
\end{equation}
In the case of a continuous frequency distribution of photons, one must replace the sum by an integral :
\begin{equation}
\langle e_1e_2|U^{(2)}|g_1g_2\rangle =e^{-it(\omega_1+\omega_2+H_f^0/\hbar)}\frac{L^2}{4 \pi^2 c^2}\iint d\omega_m d\omega_n a(\omega_m) a(\omega_n) \mathcal{A}_{mn}  \label{cont}
\end{equation}
 where $a(\omega)$ in the annihilation operator of a monochromatic photon of frequency $\omega$.

Note that the coefficient $\mathcal{A}_{mn}$ is the product of two factors which represent the response of each atom to the field. When time $t$ goes to infinity these two factors behave roughly like Dirac delta functions centered on individual atom resonances (we will precise this argument in section VI). The 2P2A excitation probability is indeed induced by the wings of the incident light spectrum which are resonant with the atoms. Consequently, if one photon is absorbed by one atom, there is no reason why the second photon should be absorbed by the second atom in a way correlated to the absorption of the first photon. In other words, the 2P2A excitation phenomenon has no reason a priori to have a resonant behavior when the 2P2A resonance condition $\omega_1+\omega_2 = \omega_m+ \omega_n$ is fulfilled.

However, as the process is non-linear and involves two atoms, it can be enhanced by taking advantage of correlation effects between the atoms or between the photons :

\begin{itemize}

\item A first possibility consists in introducing an interaction between the two atoms. Let $\omega_u\simeq\omega_1+\omega_2$  be the maximal Bohr frequency of the two-atom system :  if one photon with frequency $\omega_k$ is absorbed, then the two atoms will be more likely to absorb another photon with frequency $\omega_u-\omega_k$  in a resonant two-photon process \cite{Jose,Andrews83}. This was experimentally demonstrated in \cite{white,Sandoghdar,bagnato}
using nearly degenerate photon pairs.

\item A second possibility is to use correlated photons to interact with the two atoms. Let us consider a source {that emits correlated photons : if a photon is absorbed by one atom, then the remaining atom will interact with its correlated photon with a higher probability, leading to enhanced 2P2A resonance (we will precise this argument and the kind of correlation needed in section VII).} \end{itemize}

We will now precise these latter ideas by having a closer look at different possible light states likely to induce such a 2P2A transition.

\section{Entangled, correlated-SEPARABLE and factorized two-photon states}

Before we go further, let us precise the different kinds of two-photon states that we will consider in the following. Starting from entangled {pure} quantum state {$|\Psi\rangle$, having a density matrix} $\rho_0{=|\Psi\rangle\langle\Psi|}$ of matrix elements $\rho_{k k' q q'}= \langle 1_k,1_q|\rho_0|1_{k'}, 1_{q'} \rangle$, one can construct 
  others that have the same mean energy and the same single photon spectrum, and hence that would give the same transition probabilities for a single photon resonance. We choose two special cases that will allow a quantitative evaluation of the role of correlations :

\begin{itemize}

\item The first one is defined as
\begin{equation}
\rho_1=\sum_{k,q}\text{Tr}[\Pi_k\otimes\Pi_q\rho_0] \Pi_k\otimes\Pi_q= \sum_{k,q} \rho_{kkqq}|1_k, 1_q\rangle \langle 1_k, 1_q|
\end{equation}
where $\Pi_k$ is the frequency projection operator $\Pi_k=|1_{k}\rangle\langle1_{k}|$. $\rho_1$ is the diagonal part of $\rho_0$. It has lost any temporal field coherence and is time independent. It is actually a \textit{{correlated-}separable state}\cite{Duan}, which results from the "disentanglement" of the previous one. It gives rise however to correlations between its two parties.

\item The second
 one is defined as \begin{equation}
\rho_2=\sum_{k} \text{Tr}[\Pi_k\rho_0] \Pi_k \otimes \sum_q \text{Tr}[\Pi_q\rho_0] \Pi_q= \sum_{k,q} \sum_{q'}\rho_{kkq'q'}|1_k \rangle \langle 1_k |  \otimes \sum_{k'}\rho_{k'k'qq}| 1_q\rangle \langle1_q |\end{equation}
{This} is a fully factorized state, which does not give rise to any correlation whatsoever.
\end{itemize}

These states will induce 2P2A excitation with respective probabilities $P_1(t)$ and $P_2(t)$.  

{The two diagonal density matrices $(\rho_1,\rho_2)$ describe two c.w. fields whilst the entangled pure state $\rho_0$ describes a pulse, as a result, at a time $t$, the flow of energy having interacted with atoms in each state is different.}  {However, as shown in appendix B, this quantity is quite the same in each state when $t=L/c$. For comparison, we will take $t=L/c$ through the whole paper}.
 
We are of course interested in cases where one observes an increase in the excitation probability, i.e. when {$P(L/c) \gg P_2(L/c)$}; if  ${P(L/c) \gg P_1(L/c)}$, then entanglement is indeed the key
to efficient 2P2A {transition}, whereas if  ${P(L/c) \simeq P_1(L/c) \gg P_2(L/c)}$, correlations, of quantum or classical origin, are more important than entanglement in the present problem.

\section{2 photon 2 atom excitation induced by different photon quantum field states}

 We will now examine the efficiency of various multimode light states for the simultaneous excitation of the two atoms. 

 \subsection{two quasi-monochromatic uncorrelated photons}

  Let us begin by the simplest case : two uncorrelated photon wavepackets of mean frequencies $\omega_{\alpha}$ and $ \omega_{\beta}$, and respective spectral widths $\gamma_{\alpha}$ and $\gamma_{\beta}$ { much bigger than the detecting atom spectral widths $\gamma_{1}$ and $\gamma_{2}$ }, emitted by two uncorrelated atoms excited at the same time in the past and arriving at the detecting atoms position at $t=0$, described therefore by the bi-photon state $|\psi^{11}(t)\rangle$ with
  \begin{equation}\label{factorized}
|\psi^{11}(t)\rangle=\sum_{kq}c^{11}_{kq}e^{i(\omega_k + \omega_q)t} |{1_k,1_q} \rangle \quad ;\quad  c^{11}_{kq} =\frac{g_{\alpha}(\omega_k)g_{\beta}(\omega_q)}{
  (\omega_k-\omega_{\alpha}+i\gamma_{\alpha})(\omega_q-\omega_{\beta}+i\gamma_{\beta})}
\end{equation}
 It is the tensor product of two single-photon wave packets\cite{FabreMMQO} of duration $\gamma_{\alpha}^{-1}$ and $\gamma_{\beta}^{-1}$. In the calculation of the probability amplitude in (\ref{proba}), we will replace the sum over modes by the double integral $(L^2/4 \pi^2 c^2)\iint_0^{\infty} d\omega_k d\omega_q$, extend each integration domain to the whole real axis and use the residue theorem \cite{ScullyQObook}. For the transition probability amplitude, one finds, when
 $\gamma_{\alpha} t \gg 1$ and $\gamma_{\beta} t \gg 1$  while keeping $\gamma_{1} t \ll 1$ and $\gamma_{2} t \ll 1$ :
\begin{equation}
A^{11} \simeq \frac{L^2 f_1(\omega_1)f_2(\omega_2) g_{\alpha}(\omega_1)g_{\beta}(\omega_2)}{c^2 ( \omega_1-\omega_{\alpha}+i\gamma_{\alpha})(\omega_2-\omega_{\beta}+i\gamma_{\beta})}
\end{equation}
If we assume that the coefficients $g$ do not vary with frequency the normalization of the two-photon state imposes that :
\begin{equation}
g_{\alpha}g_{\beta}= \frac{2 c \sqrt{\gamma_{\alpha}\gamma_{\beta}}}{L}
\end{equation}
so that the transition probability $P^{11}$ is

\begin{equation}\label{11}
P^{11} = \frac{P_0 \gamma_{\alpha}\gamma_{\beta}}{[(\omega_1-\omega_{\alpha})^2+\gamma_{\alpha}^2][(\omega_2-\omega_{\beta})^2+\gamma_{\beta}^2]}
\end{equation}
where {$P_0=d_1^2 d_2^2 \omega_1 \omega_2/\hbar^2 \varepsilon_0^2 c^2 S^2=36\pi^2\gamma_1\gamma_2c^4/\omega_1^2\omega_2^2S^2$}.

\subsubsection{double resonance}

Let us first assume that we are in the best possible situation, where the photons are separately resonant with the two atoms :  $\omega_\alpha= \omega_1$ and $\omega_\beta= \omega_2$. The transition probability is then equal to
\begin{equation}\label{res}
P^{11}_{DR}= \frac{P_0 }{\gamma_{\alpha}\gamma_{\beta}}
\end{equation}
which can be written in a more general way
\begin{equation}
P^{11}_{DR}= \frac{P_0 }{S_{fr}}
\end{equation}
where $S_{fr}$ is the effective area of frequency distribution $|c_{kq}|^2$ in the $(\omega_k,\omega_q)$  {plane} (see figures (\ref{frequencycor})). This result turns out to be general and implies that all pure states having the same effective areas $S_{fr}$, entangled or not, will produce the same doubly resonant transition probability. Thus we regard (\ref{res}) as a universal result under the double resonance condition, and its value will serve as a reference for all subsequent transition probabilities. 

{In an actual experimental situation, one may take : $\gamma_{1,2}\sim 1$kHz, $\gamma_{\alpha,\beta}\sim 1$MHz, $S\simeq4\pi^2c^2/\omega_1\omega_2$, thus $P^{11}_{DR}\simeq 9\gamma_1\gamma_2/4\pi^2\gamma_{\alpha}\gamma_{\beta}\sim 10^{-7}$.}

\subsubsection{two-photon two-atom resonance}

Let us now turn to the 2P2A resonant case,  where none of the two photons are resonant with the two atoms, but where the sum of their two energies almost matches the sum of the two atomic energies : $\omega_{\alpha} + \omega_{\beta} \simeq \omega_1 + \omega_2$. The transition probability (\ref{11}) has in this case no resonant variation as a function of the 2P2A detuning $\delta=\omega_{\alpha} + \omega_{\beta} - \omega_1 - \omega_2 $. When $\delta =0$ the transition probability is :
 \begin{equation}\label{P11}
P^{11}_{2P2A}= \frac{P_0 \gamma_{\alpha} \gamma_{\beta}}{\Delta^4}=P^{11}_{DR} \frac{\gamma_{\alpha}^2 \gamma_{\beta}^2}{\Delta^4}
\end{equation}
where $\Delta$ is the smallest frequency mismatch between the emitting atoms frequencies and the detecting atoms frequencies, supposed to be much larger than the atomic widths.  Without loss of generality, we have taken $\Delta=|\omega_{\alpha} - \omega_1| = |\omega_2 - \omega_{\beta}|$.

We then conclude that the { special case of} 2P2A excitation probability by uncorrelated photons is { also} non zero for any couple of frequencies $\omega_{\alpha}, \omega_{\beta} $,  thus such a two-photon transition turns out not to be disallowed but simply induced by the wings of the two single photon frequency resonances. It is therefore very weak, as witnessed by the $\Delta^{-4}$ variation of probability.

 \subsection{two photons produced by an atomic cascade}

 Let us now envision the case considered in \cite{scully04} of a two-photon light state produced by a three-level atom excited at a given time $t=0$ in the upper state that cascades down to the ground state on two successive transitions of Bohr frequencies successively equal to $\omega_{\alpha}$ and $\omega_{\beta}$. The corresponding spontaneous emission rates are $\gamma_{\alpha}$ and $\gamma_{\beta}$. We assume that the emitted light is wholly directed in the $Oz$ direction of atoms (1) and (2) (by means of a parabolic mirror for example). It is described by a bi-photon wave-packet with a coefficient $c_{kq}^{cas}$ equal, at a time $t$ long compared to the lifetimes of the two transitions, to{\cite{scully04,ScullyQObook}} :
  \begin{equation}\label{entangled}
    c_{kq}^{cas} =\frac{g_{\alpha}(\omega_k)g_{\beta}(\omega_q)}{
  [\omega_k+\omega_q-\omega_{\alpha}-\omega_{\beta}+i\gamma_{\alpha}][\omega_q-\omega_{\beta}+i\gamma_{\beta}]}
\end{equation}
Here this entangled non stationary state is produced by a cascade, so that the photon of frequency $\omega_q$ always arrives just after the photon of frequency $\omega_k$. In addition, the probability to have   photons  of frequency sum $\omega_k+\omega_q$ close to $\omega_{\alpha} + \omega_{\beta}$ is high. We have therefore an entangled state which is  not only correlated in time  but also  anti-correlated in frequency. It is the time-energy analog of the position-momentum entangled state introduced by EPR, or of the field quadrature entangled state\cite{ScullyQObook,Reid09,Howell}.

Using the Residue Theorem, the transition probability amplitude reads exactly
\begin{equation}\label{generalSol}
A^{cas}= \frac{L^2}{c^2} \frac{g_{\alpha}(\omega_1)g_{\beta}(\omega_2)f_{1}(\omega_1) f_{2} (\omega_2)}{ \omega_{\beta2}-\delta-i(\gamma_{\beta}-\gamma_{\alpha})}\left[\frac{1-e^{-(\gamma_{\beta}+i\omega_{\beta2})t}}{\omega_{\beta2}-i\gamma_{\beta}}-\frac{1-e^{-(\gamma_{\alpha}+i\delta)t}}{\delta-i\gamma_{\alpha}}\right]+(1\leftrightarrow2)
 \end{equation}
When $\gamma_{1,2}^{-1}\gg t\gg\gamma_{\alpha,\beta}^{-1}$, the four decaying terms in Eq.(\ref{generalSol}) are negligible,  leading to a compact expression
  \begin{equation}
A^{cas}= -\frac{L^2}{c^2}\frac{f_{1}(\omega_1) f_{2} (\omega_2)}{\delta-i\gamma_{\alpha}}\left[\frac{g_{\alpha}(\omega_1)g_{\beta}(\omega_2)}{\omega_{\beta2}-i\gamma_{\beta}} +
\frac{g_{\alpha}(\omega_2)g_{\beta}(\omega_1)}{\omega_{\beta1}-i\gamma_{\beta}} \right]
\end{equation}
where $\omega_{\mu\nu}=\omega_{\mu}-\omega_{\nu}$ is the frequency difference between frequency $\omega_{\mu}$ and frequency $\omega_{\nu}; \mu,\nu=\alpha,\beta,1,2,k,q$.

\subsubsection{double resonance}

Let us first consider here also the most favorable case, which is the double resonance (DR) situation. Keeping only the largest term, one obtains in this case for the probability amplitude when $\gamma_{\alpha} t \gg 1$ and $\gamma_{\beta} t \gg 1$ :
\begin{equation}
A^{cas} _{DR}\simeq \frac{L^2 f_1(\omega_1)f_2(\omega_2) g_{\alpha}(\omega_1)g_{\beta}(\omega_2)}{c^2 \gamma_{\alpha}\gamma_{\beta}}
\end{equation}
Using the same assumption as in the previous calculation, one finds for the probability
\begin{equation}
P^{cas}_{DR} = \frac{P_0}{ \gamma_{\alpha}\gamma_{\beta}}=P^{11}_{DR}
\end{equation}
It is time independent because we are considering times much longer than the two-photon pulse of duration $\gamma_{\alpha}^{-1}+ \gamma_{\beta}^{-1}$. As it is equal to the probability obtained with uncorrelated photons, we conclude that entanglement does not help in the fully  resonant case, but does not harm either.

\subsubsection{two-photon two-atom resonance}

Let us now turn to the 2P2A resonance case. One obtains in this case for the probability :
\begin{equation}\label{pca}
P^{cas}_{2P2A}\simeq \frac{L^2}{4c^2}\frac{P_0}{ \delta^2+\gamma^2_{\alpha}}\left[\frac{g_{\alpha}(\omega_1)g_{\beta}(\omega_2)}{\omega_2-\omega_{\beta}}
+\frac{g_{\alpha}(\omega_2)g_{\beta}(\omega_1)}{\omega_1-\omega_{\beta}}\right]^2
\end{equation}

This expression, already obtained in \cite{ScullyQObook}, shows that for this state the probability has indeed a resonant character around the two-atom two-photon resonance $\delta=0$. The transition probability $P^{cas}_{2P2A}$ at the exact two-atom two-photon resonance is then :
\begin{equation}\label{pcb}
P^{cas}_{2P2A} \simeq   \frac{P_0 }{\gamma_{\alpha}\gamma_{\beta}}\frac{\gamma_{\beta}^2}{\Delta^2}=P_{DR}^{11}\frac{\gamma_{\beta}^2}{\Delta^2}    ;
\end{equation}

One therefore finds that the transition probability is in the present case smaller than $P^{11}_{DR}$ by a factor $(\gamma_{\beta}/\Delta)^2$ at exact 2P2A resonance, as expected because one is now less resonant than in the double resonance case. One finds more importantly that $P^{casc}_{2P2A}$ \textit{ is larger than $P^{11}_{2P2A}$, i.e. than in the two uncorrelated photon case, by a factor $(\Delta/\gamma_{\alpha})^2$}, which can be very large. This enhancement of the 2P2A transition probability  is the main result of \cite{scully04} : entanglement may indeed significantly enhance the two-photon two-atom process. To the best of our knowledge no experiment has been undertaken to show such a striking effect.

It must be emphasized that the present considerations do not imply that the atom cascade entangled state is the only one likely to produce such a significant increase in the transition probability. This is the reason why we will now consider other light quantum states which may also be of interest in the present problem.

  \subsection{Correlated and factorized states analogous to the atomic cascade}

Let us now consider the two states that have the same energy and the same spectrum that we have introduced in section (IV) , namely the correlated-separable state :
\begin{equation}\label{sepcasc}
\rho_1=\left(\frac{2c}{L}\right)^2\sum_{kq}\frac{\gamma_{\beta}}{(\omega_{q\beta}^2+\gamma_{\beta}^2)}\frac{\gamma_{\alpha}}{[(\omega_{q\beta}+\omega_{k\alpha})^2+\gamma_{\alpha}^2]}|1_k,1_q\rangle\langle1_k,1_q|
\end{equation}
and the factorized state :
\begin{equation}\label{fact}
\rho_2=\left(\frac{2c}{L}\right)^2\left(\sum_k\frac{\gamma_{\alpha}+\gamma_{\beta}}{\omega_{k\alpha}^2+(\gamma_{\alpha}+\gamma_{\beta})^2}|1_k\rangle\langle1_k|\right)\otimes\left(\sum_{q}\frac{\gamma_{\beta}}{\omega_{q\beta}^2+\gamma_{\beta}^2}|1_q\rangle\langle1_q|\right)
\end{equation}
The first one corresponds to an atomic cascade for which the starting time is random, thereby averaging to zero all the off-diagonal time dependent terms in the density matrix, the second one characterizes a mixed state with two uncorrelated photons having the same spectrum than the initial cascade state. They give rise to the following transition probabilities :

\begin{alignat}{2}
P_1&\simeq  P_0  \frac{\gamma_{\alpha}\gamma_{\beta}}{\delta^2+\gamma_{\alpha}^2}\left(\frac{1}{(\omega_1 - \omega_{\beta})^2}+\frac{1}{(\omega_2- \omega_{\beta})^2}\right) \frac{t^2}{(L/c)^2} & &\\
P_2&\simeq   P_0 \gamma_{\beta}(\gamma_{\alpha}+\gamma_{\beta}) \left(\frac{1}{(\omega_1 - \omega_{\beta})^4}+\frac{1}{(\omega_2- \omega_{\beta})^4}\right) \frac{t^2}{(L/c)^2}  & &
\end{alignat}
At exact 2P2A resonance, we have  $P_1 \simeq P_{DR}^{11}\gamma_{\beta}^2c^2 t^2/(\Delta^2 L^2)$ and $P_2 \simeq P_{DR}^{11} \gamma_{\alpha}\gamma_{\beta}^2( \gamma_{\alpha}+\gamma_{\beta}) c^2 t^2/(\Delta^4 L^2)$. At any time $t$, one finds $P_1\gg P_2$, since the spectral widths are much smaller than the 2P2A detuning. This fact shows that correlations play indeed an important role in the efficiency of the excitation.

Note that $P_1$ and $P_2$ depend on time, as can be expected in a situation where the detecting atoms, which have an infinite lifetime, are submitted to a stationary quantum state,  and therefore to c.w. light. In order to compare $P_1$ and $P_2$ to $P^{cas}_{2P2A}$ (equation (\ref{pca})), which is induced by a pulse of light, we need to fix an interaction time $t$. It is shown in appendix B that the two atoms are submitted to the same energy flow at time $t=L/c$.  One then obtains at this time and at exact resonance :

\begin{equation}
P_1 \simeq P_{DR}^{11}\frac{\gamma_{\beta}^2}{\Delta^2} \simeq  P^{cas}_{2P2A};
\end{equation}
We thus find the result that \textit{a correlated-separable state like $\rho_1$ can induce the 2P2A transition as efficiently as the entangled cascade state}.  This statement constitutes the main result of the present paper.

Let us stress that $\rho_1$, though not entangled, has indeed genuine quantum properties, being a mixture of single photon states which are highly non-classical. It displays strong correlations that we will study in more detail in section VI.

\subsection{two-photon state produced by parametric down conversion}

Let us now examine the two-photon state  $|\Psi_{pdc}\rangle$  produced by non-degenerate parametric down conversion which has been under wide and in-depth investigation for many years. Because of its $\chi^{(2)}$ nonlinearity, a non-linear crystal submitted to a pulsed pump field of central frequency $\omega_{\alpha}+\omega_{\beta}$ and narrow bandwidth $\sigma_{\alpha}$ emits a signal field (central frequency $\omega_{\alpha}$) and an idler field (central frequency $\omega_{\beta}$). Let  $\sigma_{\beta}$ be the frequency width of the phase matching curve. For the sake of computational simplicity we will use a Gaussian approximation for both the laser lineshape and the phase matching curve. The crystal generates in such a case an entangled state which is described by a wavepacket with a coefficient $c_{kq}^{pdc}$ \cite{wang2006} given by
\begin{equation}\label{SPDC}
c_{kq}^{pdc}=\mathcal{N}e^{-\frac{(\omega_{k\alpha}+\omega_{q\beta})^2}{2 \sigma_{\alpha}^2}+i(\omega_{k\alpha}+\omega_{q\beta})t_0}
\left(e^{-\frac{\omega_{k\alpha}^2+\omega_{q\beta}^2}{2 \sigma_{\beta}^2}}+ ie^{-\frac{\omega_{k\beta}^2+\omega_{q\alpha}^2}{2 \sigma_{\beta}^2}}
\right)
\end{equation}
where $\mathcal{N}$ is normalized coefficient, satisfying
$$\left(\frac{L}{2\pi c}\right)^2\mathcal{N}^2\frac{2\pi \sigma_{\alpha} \sigma_{\beta}^2}{\sqrt{ \sigma_{\alpha}^2+2 \sigma_{\beta}^2}}=1$$
In the expression of (\ref{SPDC}), we have assumed that the pump laser pulse had a Gaussian temporal shape centered at time $t_0 \gg  \sigma_{\alpha}^{-1}+\sigma_{\beta}^{-1}$ to provide most of the photons a chance to interact with the two detecting atoms. The factor $i$ in the second component originates from a relative phase (depends on the birefringence) which is set to be $\pi/2$ for the sake of simplicity in our case.

Here we will also extend the double integral to the whole plane and find, when $t$ is sufficient large\cite{handbook},
 the transition probability
\begin{equation}
P^{pdc}=\pi P_0\frac{\sqrt{ \sigma_{\alpha}^2+2 \sigma_{\beta}^2}}{ \sigma_{\alpha} \sigma_{\beta}^2}e^{-\frac{\delta^2}{ \sigma_{\alpha}^2}}\left(e^{-\frac{\omega_{1\alpha}^2+\omega_{2\beta}^2}{2 \sigma_{\beta}^2}}+
e^{-\frac{\omega_{2\alpha}^2+\omega_{1\beta}^2}{2 \sigma_{\beta}^2}}
\right)^2
\end{equation}

Let us also take into account the two mixed biphoton states ($\rho_{1}^{pdc}, \rho_{2}^{pdc}$) pertaining to the pure SPDC type II biphoton state (\ref{SPDC}),
\begin{alignat}{2}
&\rho_{1}^{pdc}=\mathcal{N}^2\sum_{kq}e^{-\frac{(\omega_{k\alpha}+\omega_{q\beta})^2}{ \sigma_{\alpha}^2}}\left(e^{-\frac{\omega_{k\alpha}^2+\omega_{q\beta}^2}{ \sigma_{\beta}^2}}+e^{-\frac{\omega_{k\beta}^2+\omega_{q\alpha}^2}{ \sigma_{\beta}^2}}
\right)|1_k,1_q\rangle\langle1_k,1_q|,\label{sepSPDC} & &\\
&\rho_{2}^{pdc}=\pi\frac{c^2}{L^2}\frac{\zeta}{ \sigma_{\beta}^2}\left[\sum_{k}\!\!
\left(
e^{-\zeta\frac{\omega_{k\alpha}^2}{ \sigma_{\beta}^2}}+
e^{-\zeta\frac{\omega_{k\beta}^2}{ \sigma_{\beta}^2}}\right)|1_k\rangle\langle1_k|\right]\!\!\otimes\!\!\left[\sum_q\!\!
\left(
e^{-\zeta\frac{\omega_{q\alpha}^2}{ \sigma_{\beta}^2}}+
e^{-\zeta\frac{\omega_{q\beta}^2}{ \sigma_{\beta}^2}}\right)|1_q\rangle\langle1_q|\right],& &
\end{alignat}
where $\zeta=1+ \sigma_{\beta}^2/( \sigma_{\alpha}^2+\sigma_{\beta}^2)$. The first one corresponds to a SPDC process in which all the off-diagonal time dependent terms in the density matrix  are averaging to zero by random processes, while the second one characterizes a mixed state with two uncorrelated photons having the same spectrum than the initial SPDC state. When $t$ is sufficient large, their corresponding transition probabilities read\begin{alignat}{2}
P_1^{pdc}&=\pi P_0
\frac{\sqrt{ \sigma_{\alpha}^2+2 \sigma_{\beta}^2}}{ \sigma_{\alpha} \sigma_{\beta}^2}
e^{-\frac{\delta^2}{ \sigma_{\alpha}^2}}\left(e^{-\frac{\omega_{1\alpha}^2+\omega_{2\beta}^2}{ \sigma_{\beta}^2}}+e^{-\frac{\omega_{1\beta}^2+\omega_{2\alpha}^2}{ \sigma_{\beta}^2}}
\right) \left(\frac{t}{L/c}\right)^2 &&\\
P_2^{pdc}&=\pi \frac{P_0}{2}\frac{\zeta}{ \sigma_{\beta}^2}\left(
e^{-\zeta\frac{\omega_{1\alpha}^2}{ \sigma_{\beta}^2}}+
e^{-\zeta\frac{\omega_{1\beta}^2}{ \sigma_{\beta}^2}}\right)
\left(
e^{-\zeta\frac{\omega_{2\alpha}^2}{ \sigma_{\beta}^2}}+
e^{-\zeta\frac{\omega_{2\beta}^2}{ \sigma_{\beta}^2}}\right) \left(\frac{t}{L/c}\right)^2&&
\end{alignat}
 We will once again take $t=L/c$ to be able to compare in a fair way the pulsed and c.w. excitations through the whole following discussions.

 \subsubsection{double resonance}

Let us first consider the DR situation with $\omega_{\alpha}=\omega_1$ and $\omega_{\beta}=\omega_2$. Keeping the largest term, one finds the probability
\begin{equation}
P^{pdc}_{DR}=P_{1,DR}^{pdc}\simeq\pi P_0\frac{\sqrt{ \sigma_{\alpha}^2+2 \sigma_{\beta}^2}}{ \sigma_{\alpha} \sigma_{\beta}^2};
\end{equation}
Once again, we conclude that entanglement is not active in enhancing the transition probability in the double resonance case.

One also finds $P_{DR}^{pdc}\simeq P_{DR}^{11}$ when $\sigma_{\alpha}=\gamma_{\alpha}, \sigma_{\beta}=\gamma_{\beta}$. In the following we will take this correspondences of spectral widths for comparisons. Henceforth, $P_{DR}^{pdc}$ or $P_{DR}^{11}$  will be regarded as a reference in the discussions related to SPDC biphoton state.

\subsubsection{two-photon two-atom resonance}

Now we will turn to the 2P2A case. The transition probability $P_{2P2A}^{pdc}$ has indeed a resonant character around $\delta=0$. At the exact 2P2A resonance, it is equal to
\begin{equation}
P^{pdc}_{2P2A}\simeq P_{DR}^{11}e^{-2\Delta^2/ \sigma_{\beta}^2}
\end{equation}
which is much smaller than for the atom cascade state because the factor $\Delta^2/ \sigma_{\beta}^2$ enters now as exponent in a Gaussian function and the detuning $\Delta$ is much greater than the spectral widths.

For the factorized, uncorrelated state $\rho_2^{pdc}$, the transition probability in this case reads
\begin{equation}
P_{2,2P2A}^{pdc}\simeq P_{DR}^{11}(1+2\sigma_{\beta}^2/\sigma_{\alpha}^2)^{-1/2}e^{-2\zeta\Delta^2/ \sigma_{\beta}^2}
\lesssim e^{-2\Delta^2/( \sigma_{\alpha}^2+ \sigma_{\beta}^2)}P^{pdc}_{2P2A}
 \end{equation}
Thus, $P^{pdc}_{2P2A}$ is  much greater than the probability given by the factorized state because of the scale factor $e^{2\Delta^2/( \sigma_{\alpha}^2+ \sigma_{\beta}^2)}$. So we obtain in the Parametric Down Conversion configuration the same conclusion as the one drawn in \cite{scully04} for the atomic cascade : the entangled state  $|\Psi_{pdc}\rangle$ is much more efficient for inducing a 2P2A resonance than the factorized, uncorrelated state.

For the correlated-separable state $\rho_1^{pdc}$, the transition probability reads
\begin{alignat}{2}
P_{1,2P2A}^{pdc}&\simeq P_{DR}^{11}e^{-2\Delta^2/ \sigma_{\beta}^2} \simeq P_{2P2A}^{pdc}
\end{alignat}
Thus, one has $P_{1,2P2A}^{pdc}\gg P_{2,2P2A}^{pdc}$. The same conclusion is found as in the cascade case : the correlated-separable  state is as efficient as the entangled state to boost the 2P2A resonance. The fact that $P_{1,2P2A}^{pdc}$ is much larger than $P_{2,2P2A}^{pdc}$ and $P_{1,2P2A}^{pdc}\simeq P_{2P2A}^{pdc}$ once again shows that correlations, which are not necessarily related to entanglement, play indeed a crucial role in the efficiency of the excitation.

\section{Enhancement of 2P2A resonance for more general classes of light states}

We have so far studied interesting but specific states of light and showed an enhancement effect for some of them, entangled or correlated-separable. It would be interesting to consider now more general classes of light states.

\subsection{Light pulses starting at a given time}

Let us go back to the initial equations (\ref{ampl}) and  (\ref{proba}). They contain functions like $[1-\exp(i\omega_{1m}t)]/\omega_{1m}$. As explained in the appendix A, even though this function does not act as a Dirac function when it is applied to integrations with any function, it indeed tends to $ 2 i \pi \delta (\omega_1-\omega_m)$ when $t \rightarrow \infty$ when applied to functions of $\omega_m$ that have a Fourier transform which is strictly zero for $t<0$. Such will be the case here.

The initial two-photon light state $|\Psi \rangle$ is the pure state :
\begin{equation}\label{initial}
	|\Psi\rangle=\sum_{kq}   c_{kq}|1_k,1_q\rangle,
\end{equation}
  Let us assume that this state describes a "switched-on" light which is not vacuum only after time $t=0$. One can then use the delta function approximation. The probability that the two atoms are found in the excited state at times long compared to the pulse duration is  now
\begin{equation}\label{generalP}
	P\simeq \frac{P_0}{4}\frac{L^2}{c^2}
	|c_{12}+c_{21}|^2.
\end{equation}
Mathematically, if  $|c_{12}|\sim|c_{21}|$ this interference, which has been
studied in the literature \cite{Fei}, may lead to strong
variations according to the relative phase. According to the Cauchy-Schwatz inequality, one has
\begin{equation}
0\leq P\leq 2\left(\frac{P_0}{4}\frac{L^2}{c^2}(|c_{12}|^2+|c_{21}|^2)\right).
\end{equation}
 However, physically speaking, only one component between $c_{12}$ and $c_{21}$ dominates in the expression (\ref{generalP}). This is because  we have assumed that the quantities $\omega_1, \omega_2, \omega_{\alpha}, \omega_{\beta}$ are sufficiently separated from each other but with a small 2P2A detuning $\delta\simeq0$, as a result, $\omega_1$ should be closer to one of the central frequencies of the fields than to the rest one. Under this condition, one has
 \begin{equation}
 P\simeq \frac{P_0}{4}\frac{L^2}{c^2}(|c_{12}|^2+|c_{21}|^2).
\end{equation}
The correlated and factorized states $\rho_1, \rho_2$ analogous to the initial state $|\Psi\rangle\langle\Psi|$, give rise to the following 2P2A transition probabilities :
\begin{alignat}{2}
&P_1=\frac{P_0}{4}t^2\left(|c_{12}|^2+|c_{21}|^2\right) \label{prob1}   &&\\
&P_2= \frac{P_0}{4}t^2\sum_{mn}\left(|c_{1n} c_{m2}|^2+|c_{2n} c_{m1}|^2\right) \label{prob2}&&
\end{alignat}

A discriminability index in the role of enhancing 2P2A transition probability is defined by the quotient $G_p$ between $P$ and $P_1$ at $t=L/c$,
\begin{equation}
G_p=\left. \frac{P}{P_1}\right|_{t=L/c}=\frac{|c_{12}+c_{21}|^2}{|c_{12}|^2+|c_{21}|^2}
\end{equation}
Thus, one finds $0\leq G_p\leq2$. 
The maximum value 2 is achieved when $c_{12}=c_{21}$.

One has $G_p\simeq1$ under the physical conditions we stated before. That is, the entangled and the correlated-separable state yield almost equal transition probabilities. This implies that the conclusion that we had drawn in the special previous cases is valid for a large class of two-photon states : correlated states are as efficient as entangled states in 2P2A co-excitation when they have delivered the same amount of energy to the two atoms.

Another important discriminability index is the ratio between the two transition rates $P_1$ and $P_2$ :
\begin{equation}\label{freq}
G_{12} =\frac{|c_{12}|^2+|c_{21}|^2}{ \sum_{mn} \left(|c_{1n} c_{m2}|^2+|c_{2n}c_{m1}|^2\right)}
\end{equation}
The value of the enhancement factor $G_{12}$ can be used as a witness for
the correlation needed in such a problem.

Note in addition that, while $P$ is sensitive to possible destructive interference effects between $c_{12}$ and $c_{21}$,  $P_1$ is not. Therefore, the enhancement
effect as indicated by G, and due to correlations not related to
entanglement,  turns out to be more "robust" than the one related to it.

\subsection{Coherent states}

So far we have only considered biphoton states of different shapes, which are all strongly non-classical objects, as they are produced by spontaneous emission or parametric fluorescence which are specifically quantum processes with no classical equivalent. But one can also envision superpositions of two-mode coherent states of the form :

\begin{equation}\label{coherent}
	|\Psi_{coh}\rangle=\sum_{kq} c_{kq} 	|\alpha (\omega_k)\rangle \otimes |\alpha(\omega_q)\rangle,
\end{equation}
where $|\alpha(\omega_k)\rangle$ is the coherent state $|\alpha\rangle$ in the mode of frequency $\omega_k$, $\alpha$ being the same complex number for all modes.

The calculation of the transition probability must be redone from the beginning. By using the approximation $\langle\alpha|0\rangle\approx0$ valid for $|\alpha|\gg1$, one finally finds :
\begin{equation}\label{coherent2}
	P_{coh}(t) =|\alpha|^4 P(t)
\end{equation}
where $P(t)$ is the probability (\ref{proba}) obtained for two-photon states. Apart from the energy scaling factor  $|\alpha|^4 $, the conclusions of the previous paragraphs hold in the present case, which looks much more classical  than the previously studied ones, as such states can be produced by classical means.

\section{What kind of correlation is required to enhance the 2P2A transition probability ?}

We have found in the previous sections that the 2P2A transition probability depends crucially on the specific state of light used for the excitation, even when all the considered states have the same energy spectrum. The question we address now is the physical origin of an enhanced transition probability. We have seen that entangled and not entangled states may give comparable results, so a first answer to the question is obviously that entanglement is not at the origin of the effect, but rather some kind of correlation effect which is shared by entangled and not entangled states.

Candidates likely to play a role in the present problem is time correlation and frequency correlation. We will now examine them successively

\subsection{Temporal correlation effect}

It is well characterized by the cross second order correlation function
$g_{\times}^{2}(t,\tau)$
\begin{equation}
g_{\times}^{(2)}(t,\tau)=\frac{\text{Tr}[\rho_0\hat{E}_{\alpha}^{(-)}(\tau)\hat{E}_{\beta}^{(-)}(t)\hat{E}^{(+)}_{\beta}(t)\hat{E}_{\alpha}^{(+)}(\tau)]}{\text{Tr}[\rho_0\hat{E}_{\alpha}^{(-)}(t)\hat{E}^{(+)}_{\alpha,}(t)]\text{Tr}[\rho_0\hat{E}_{\beta}^{(-)}(t)\hat{E}^{(+)}_{\beta}(t)]}
\end{equation}
Assuming that the amplitude of the single-photon electric field  is a smooth function of $\omega_k$, one gets for the pure state $|\Psi\rangle=\sum_{kq} c_{k,q}|1_k,1_q\rangle$,
\begin{equation}
g_{\times}^{(2)}(t,\tau)=\left|\sum_{kq}  c_{k,q} e^{-i\omega_k\tau-i\omega_qt}\right|^2
\end{equation}
It is the two-time Fourier transform of the two-photon state.

\begin{enumerate}
\item  In the case of the cascade state (\ref{entangled})
\begin{equation}
g_{\times}^{(2)}(t,\tau)=\left(\frac{L}{2\pi c}\right)^2\frac{ \gamma_{\alpha}\gamma_{\beta}}{\pi^2}\theta(\tau)\theta(t-\tau)e^{-2 \gamma_{\alpha}\tau-2 \gamma_{\beta}(t-\tau)}
\end{equation}

$\theta(t)$ being the step function. We notice here a time asymmetry between $t$ and $\tau$, expected in the case of a cascade in which the $\omega_{\alpha}$ photon is always emitted before the  $\omega_{\beta}$ photon.

\item  For the SPDC state (\ref{SPDC}) :
\begin{equation}\!\!\!\!
g_{\times}^{(2)}(t,\tau)=\frac{2}{\mathcal{N}^2}[1+\sin\omega_{\alpha\beta}(t-\tau)]\exp\!\left[-\frac{ \sigma_{\beta}^2(t-\tau)^2}{2}-\frac{2 \sigma_{\alpha}^2 \sigma_{\beta}^2}{ \sigma_{\alpha}^2+2 \sigma_{\beta}^2}\left(t_0-\frac{t+\tau}{2}\right)^2\right]
\end{equation}

\end{enumerate}

\begin{figure}
\centering
\includegraphics[scale=0.55]{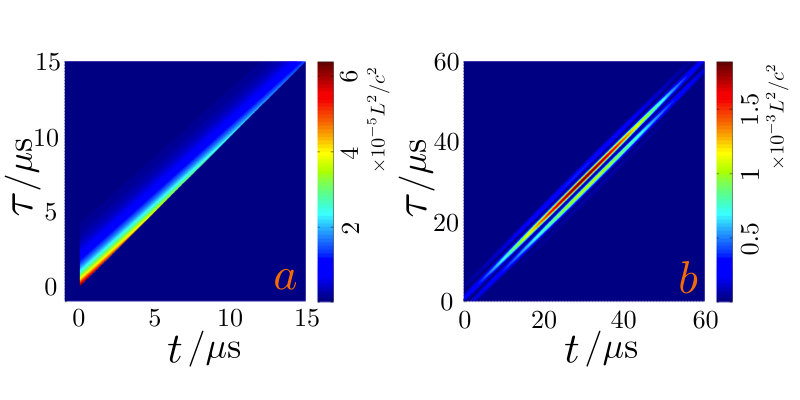}
\caption{\footnotesize Plots of the cross temporal correlation function $g_{\times}^{(2)}(t,\tau)$. The left one is for the atom cascade biphoton state, with $\gamma_{\alpha}=0.05$MHz and $\gamma_{\beta}=0.5$MHz;  the right plot is for the SPDC biphoton state, in which the pulse takes place around $t_0=30\mu$s, and with $\sigma_{\alpha}=0.05$MHz, $\sigma_{\beta}=0.5\text{MHz}, \omega_{\beta\alpha}=2$MHz. Note the ($t,\tau$) asymmetry in the first figure, and fringes in the second one due to interferences from two temporal processes. In both plots, one finds significant temporal correlations along the diagonal line. In a real condition, the value of $\omega_{\beta\alpha}$ should be much greater, leading to a poorer graphic representation for interference patterns}\label{timecor}
\end{figure}

As can be seen on the figure (\ref{timecor}), $g_{\times}^{2}(t,\tau)$ is in both cases significant only very close to the diagonal, which implies that both states exhibit strong temporal
correlations, as expected. The width of the diagonal, which gives the characteristic time of
this correlation, is equal to $\gamma_{\beta}^{-1}( \sigma_{\beta}^{-1})$ in both the cascade and SPDC cases.

It is easy to see that for the correlated-separable states (\ref{sepcasc}) and (\ref{sepSPDC}), there is no time dependence for
$g_{\times}^{(2)}(t,\tau)$,  and hence no temporal correlation, as expected from a c.w. time averaged state in which the photons arrive at any time. It is also the case for the coherent states (\ref{coherent}). As these states give 2P2A transition probabilities comparable to the entangled state, we must conclude that the temporal correlation is not the physical origin of the enhancement effect, nor the time ordering of the photons present in the cascade state. The physical reason is that, as we have neglected their spontaneous emission, the two detecting atoms have an infinite memory time, and hence they can be excited separately at any time.
\subsection{Frequency correlation effect}

It is well characterized by the cross second order frequency correlation function
$g_{\times}^{2}(\omega,\omega')$
\begin{equation}
g_{\times}^{(2)}(\omega,\omega')=\frac{\text{Tr}[\rho_0\hat{E}_{\alpha}^{(-)}(\omega')\hat{E}_{\beta}^{(-)}(\omega)\hat{E}^{(+)}_{\beta}(\omega)\hat{E}_{\alpha}^{(+)}(\omega')]}{\text{Tr}[\rho_0\hat{E}_{\alpha}^{(-)}(\omega')\hat{E}^{(+)}_{\alpha}(\omega')]\text{Tr}[\rho_0\hat{E}_{\beta}^{(-)}(\omega)\hat{E}^{(+)}_{\beta}(\omega)]}
\end{equation}
equal in the pure state case to $|c(\omega, \omega')|^2$ and to $\rho(\omega, \omega')$ in the mixed state case.

\begin{figure}
\centering
\includegraphics[scale=.6]{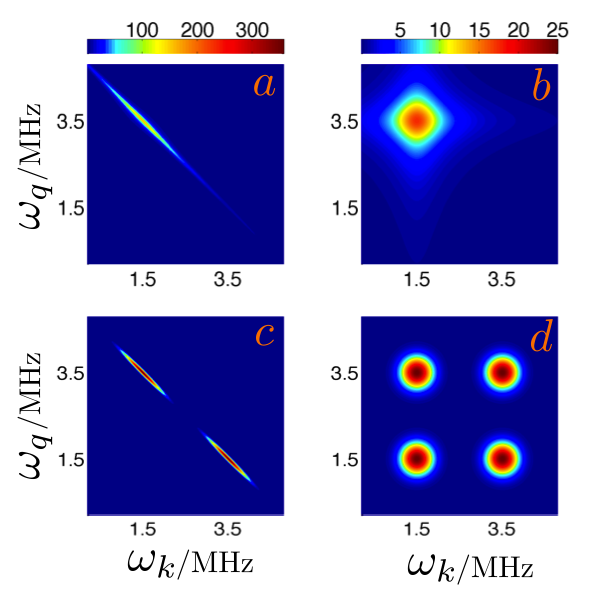}
\caption{ Plots of the cross frequency correlation function $g_{\times}^{(2)}(\omega_k,\omega_q)$. a : entangled, correlated-separable and coherent cascade states ; b : factorized cascade state; c : entangled, correlated-separable and coherent SPDC states ; d : factorized SPDC state. In all plots $\gamma_{\alpha}=\sigma_{\alpha}=0.05\text{MHz},\gamma_{\beta}=\sigma_{\beta}=0.5$MHz, $\omega_{\alpha}=1.5\text{MHz},\omega_{\beta}=3.5\text{MHz}$. The color codes, in the unit of $c^2/L^2$, on the top left (right) are shared by a and c (b and d). The left side plots exhibit strong frequency anticorrelations along the line $\omega_k+\omega_q=\omega_{\alpha}+\omega_{\beta}$, whilst in the right side plots, one finds no such a correlation.
The type II SPDC biphoton source is non-degenerate and each photon has two distribution peaks, thus one sees two bright spots in the left side bottom plot and 4 bright spots in the factorized case in the right side bottom plot. In a real condition, the distances of the peaks in the bottom side plots are much greater, and the sizes of spots are much smaller.}\label{frequencycor}
\end{figure}

 This quantity is plotted in figure (\ref{frequencycor}) for the cascade and SPDC states, either entangled, correlated-separable or factorized. One observes that the frequency correlation functions take significant values only on the anti-diagonal for the left side plots, which implies that the corresponding states exhibit  strong frequency anticorrelations. This is not the case for the right side plots. The width of the anti-diagonal, which gives the characteristic width of the frequency anticorrelation, is equal to $\gamma_{\alpha}(\sigma_{\alpha})$ in both the cascade and SPDC entangled and {correlated-}separable cases.

 Let us note that the entangled cascade and SPDC states are the only ones in our list exhibiting simultaneously time correlations and frequency anti-correlations : one has in these states EPR-like correlations, revealed by a violation of the time-energy Heisenberg inequality\cite{Howell,Barak} when $\gamma_{\alpha}/\gamma_{\beta}(\text{or}\ \sigma_{\alpha}/\sigma_{\beta})\ll1$.

The important point to notice is that such a frequency anti-correlation exists for all the states which exhibit 2P2A resonance enhancement, and is not present for the states which  do not give rise to this effect. We are therefore led to the conclusion they \textit{the property needed to enhance the 2P2A excitation is precisely the presence of strong frequency anticorrelations  in the quantum state}.

This conclusion, that we have demonstrated for the two specific examples that we have considered in the first sections of this paper, is far more general, as can be seen on the expression of the probability written for any switched on two-photon state.  

Equations (37), (38) and (39) show indeed that the probability of 2P2A excitation is proportional to the component of the density matrix of the two-photon state corresponding to the existence of one photon with frequency $\omega_1$ and one photon with frequency $\omega_2$. This gives a simple interpretation of the problem : there is 2P2A excitation only when each photon of the two-photon state is resonant with the atomic transition of the atom it excites. This is expected, since we are considering that the atomic excited states have a very long lifetime, and therefore very narrow linewidths. Since the spectrum of each photon of the source has a much larger bandwidth, the probability of excitation is small. If the photons are not correlated in frequency, the probability of double excitation is proportional to the product of the probabilities that each photon has the corresponding transition frequency, and this yields a very small transition probability. But when the photons are anti-correlated in frequency such that the sum of their frequencies is equal to the sum of the transition frequencies of the atoms, when one photon is resonant with one atomic transition, the correlated photon will be automatically resonant with the other transition, and the probability of 2P2A transition will in general be much higher than in the non-correlated case.

 We can say that the 2P2A transition occurs with a higher probability when the sum of the photon frequencies is found inside a small interval around the sum of the atomic transition frequencies, so that the enhancement is associated with the inverse of the variance of the $|c_{kq}|^2$ distribution in the direction of the diagonal.

\section{Conclusion}

We have now elements of answer to the question raised in the title and in the introduction about the role of entanglement in the two-photon excitation process considered in this paper : We have shown that what is necessary for the enhancement of the transition probability is not precisely quantum entanglement nor temporal correlations, but frequency anticorrelation, which can be due to the presence of entanglement in the state, but also to correlations that are not related to entanglement.
  
 As any nonlinear process, like two-photon absorption in a single atom \cite{Loudon}, 2P2A transition probability can be modified by changing the quantum state of light, and therefore the enhancement effect  that we have studied  in this paper is due to the partial optimization of the quantum state.

We have not treated in this paper the important question of characterizing in a quantitative way the frequency correlation  relevant to the present enhancement and relating it to its classical or quantum character though various quantum correlation witnesses such as the quantum discord. It will be addressed in a subsequent paper, together with the important question of the full optimization of the quantum state with respect to the 2P2A probability maximization, given a constant spectral energy distribution.

\textbf{Acknowledgements}

The authors thank M. Scully, H. Eleuch, G.S. Agarwal, B. Dayan and S. Mukamel for valuable discussions on the present subject. C. F. is a member of the Institut Universitaire de France. P.L.S. was supported by the
Brazilian agencies CNPQ and FACEPE; J. R. R. L. acknowledges support from Brazilian Capes-Cofecub-456 and CNPQ-Facepe-Pronex APQ0630-1.05/06. Z. Z. acknowledges China Scholarship Council for the support.

\section*{APPENDIX A : \\ Is $(\exp(-i \omega t) - 1)/(2 i \pi \omega)$ \\ a good approximation of the delta function?}

Let us note $s_t(\omega)$ the function $(\exp( - i \omega t) - 1)/(2 i \pi \omega)$. One can also write it as $s_t(\omega) = - \sin \omega t/(2 \pi \omega) + i (1-\cos\omega t)/(2 \pi \omega)$. Whereas the real part of $s_t(\omega)$  is a sinc function which tends indeed to a delta function when $t \rightarrow \infty$,  the imaginary part, being not a peaked function whose area is constant, is not an approximation of the delta function. So in general  $s_t(\omega)$ does not tend to the delta distribution when it acts on the general set of integrable functions. However, it can be so on a smaller set of functions. This set includes for example all the odd functions in $\omega$, a subset which is not relevant for the present paper. We show in this appendix that  $s_t(\omega)$ behaves also as a delta function when it acts on functions which have a Fourier transform which is strictly zero before $t=0$.

Let us consider a function $F(t)$ that is zero for $t<0$ and admits a well-behaved Fourier transform $f(\omega)$. Then
\begin{alignat}{2}
f(\omega)&=\frac{1}{2\pi}\int_{-\infty}^{\infty}\hbox{d}t F(t)e^{i\omega t}=\frac{1}{2\pi}\int_{0}^{\infty}\hbox{d}t F(t)e^{i\omega t} & &\\
F(t)&=\int_{-\infty}^{\infty}\hbox{d}\omega f(\omega)e^{-i\omega t} & &
\end{alignat}
where $f(\omega)$ is absolutely integrable, which excludes functions like $1/(\omega+i\gamma)$ from the present discussion. Let us now calculate the integral
\begin{equation}
I=\int_{-\infty}^{\infty}\hbox{d}\omega\frac{\exp(-i\omega t)-1}{\omega}f(\omega)
=i\int_0^t\hbox{d}\tau\int_{-\infty}^{\infty}\hbox{d}\omega f(\omega)\exp(-i\omega \tau)
=i\int_0^t\hbox{d}\tau\theta(\tau)F(\tau)
\end{equation}

Then $I \rightarrow i\int_0^{\infty} d\tau F(\tau)= 2\pi if(0)$ when $t\rightarrow \infty$. This proves that $s_t(\omega)$ acts as a delta function for the set of functions that have a Fourier transform strictly null for $t < 0$.

\section*{APPENDIX B : \\ why do we take $t=L/c$ in the comparison of transition probabilities ?}
In order to compare the probabilities of transitions induced by pulsed and c.w. light in a fair way, we must be careful to take the same amount of energy flow $\mathcal{F}(t$) on the detecting atoms in both cases. This quantity is nothing else than the integral over time and transverse section $S$ of the Poynting vector. It is equal to, at a given time $t$ and for a state $\rho$ :
\begin{equation}
\mathcal{F}(t)=2\varepsilon_0 c S \int_0^t \text{Tr}[\rho \hat{E}^{+\dag}(\tau)\hat{E}^+(\tau)] d\tau \simeq \hbar \omega \frac cL \int_0^t  \text{Tr}[\rho\hat{b}^{\dag}(\tau) \hat{b}(\tau)] d\tau
\end{equation}
where $\hat{b}(\tau)=\sum_m\hat{a}_m\exp(-i\omega_m\tau)$ and $\omega$ is the mean frequency of the state under consideration.

For any diagonal density matrix(DDM), since $\text{Tr}[\rho_{DDM} \hat{b}^{\dag}(\tau)\hat{b}(\tau)]=2$ is time-independent, one finds a linear relationship between the energy flow and time $t$
\begin{equation}\label{ddm}
\mathcal{F}_{DDM}(t)=2\hbar \omega  \frac {c t} L
\end{equation}

For any entangled pure state $|\Psi\rangle=\sum_{kq}c_{kq}|1_k,1_q\rangle$ :
\begin{equation}
\text{Tr}[|\Psi\rangle\langle\Psi| \hat{b}^{\dag}(\tau)\hat{b}(\tau)]=\sum_{k}|\sum_qc_{kq}e^{-i\omega_q\tau}|^2+\sum_{q}|\sum_kc_{kq}e^{-i\omega_k\tau}|^2
\end{equation}
The energy flow at time $t$ is
\begin{equation}
\mathcal{F}_{\Psi}(t)= \int_0^td\tau\text{Tr}[|\Psi\rangle\langle\Psi| \hat{b}^{\dag}(\tau)\hat{b}(\tau)]\approx  \int_{-\infty}^td\tau\text{Tr}[|\Psi\rangle\langle\Psi| \hat{b}^{\dag}(\tau)\hat{b}(\tau)]
\end{equation} 
when most photons arrive at the detecting atoms after $t=0$. One assumes that at sufficient large time $t$ (much greater than the temporal coherence length of the field), the photons in state $|\Psi\rangle$ have fully interacted with the detecting atoms, therefore, one extends $t$ to $+\infty$ without introducing notable error. By using the Parseval identity, one has
\begin{alignat}{2}
\mathcal{F}_{\Psi}(t) &\approx \hbar \omega  \frac {c} L
 \int_{-\infty}^{\infty}d\tau\text{Tr}[|\Psi\rangle\langle\Psi| \hat{b}^{\dag}(\tau)\hat{b}(\tau)]& &\notag\\
&= \hbar \omega  \frac {c} L
 \int_{-\infty}^{\infty}d\tau \left[\sum_{k}|\sum_qc_{k,q}e^{-i\omega_q\tau}|^2+\sum_{q}|\sum_kc_{k,q}e^{-i\omega_k\tau}|^2
\right]& &\notag\\
&= \hbar \omega \left[\sum_{kq}|c_{kq}|^2+\sum_{kq}|c_{kq}|^2
\right]= 2\hbar \omega
 & &
\end{alignat}
as expected. By comparison with Eq.(\ref{ddm}), one finds that at time $t=L/c$, the energies supplied by the c.w. field and by the pulse are equal. Under this situation, one can make reasonable comparisons between the corresponding transition probabilities.


\begin{thebibliography}{0}%
\makeatletter
\providecommand \@ifxundefined [1]{%
 \@ifx{#1\undefined}
}%
\providecommand \@ifnum [1]{%
 \ifnum #1\expandafter \@firstoftwo
 \else \expandafter \@secondoftwo
 \fi
}%
\providecommand \@ifx [1]{%
 \ifx #1\expandafter \@firstoftwo
 \else \expandafter \@secondoftwo
 \fi
}%
\providecommand \natexlab [1]{#1}%
\providecommand \enquote  [1]{``#1''}%
\providecommand \bibnamefont  [1]{#1}%
\providecommand \bibfnamefont [1]{#1}%
\providecommand \citenamefont [1]{#1}%
\providecommand \href@noop [0]{\@secondoftwo}%
\providecommand \href [0]{\begingroup \@sanitize@url \@href}%
\providecommand \@href[1]{\@@startlink{#1}\@@href}%
\providecommand \@@href[1]{\endgroup#1\@@endlink}%
\providecommand \@sanitize@url [0]{\catcode `\\12\catcode `\$12\catcode
  `\&12\catcode `\#12\catcode `\^12\catcode `\_12\catcode `\%12\relax}%
\providecommand \@@startlink[1]{}%
\providecommand \@@endlink[0]{}%
\providecommand \url  [0]{\begingroup\@sanitize@url \@url }%
\providecommand \@url [1]{\endgroup\@href {#1}{\urlprefix }}%
\providecommand \urlprefix  [0]{URL }%
\providecommand \Eprint [0]{\href }%
\providecommand \doibase [0]{http://dx.doi.org/}%
\providecommand \selectlanguage [0]{\@gobble}%
\providecommand \bibinfo  [0]{\@secondoftwo}%
\providecommand \bibfield  [0]{\@secondoftwo}%
\providecommand \translation [1]{[#1]}%
\providecommand \BibitemOpen [0]{}%
\providecommand \bibitemStop [0]{}%
\providecommand \bibitemNoStop [0]{.\EOS\space}%
\providecommand \EOS [0]{\spacefactor3000\relax}%
\providecommand \BibitemShut  [1]{\csname bibitem#1\endcsname}%
\let\auto@bib@innerbib\@empty
\end{thebibliography}%


\begin{thebibliography}{99}

\bibitem{Plenio07}
M. B. Plenio and S.Virmani, Quant. Inf. Comp. \textbf{7}, 1 (2007)

\bibitem{Gerardo07} G. Adesso and F. Illuminati,
J. Phys. A: Math. Theor. \textbf{40}, 7821, (2007)

\bibitem{Reid09} M. D. Reid, P. D. Drummond, W. P. Bowen, E. G. Cavalcanti, P. K. Lam and H. A. Bachor, U. L. Andersen, G. Leuchs, Rev. Mod. Phys. \textbf{81}, 1727 (2009)


\bibitem{Ryszard09}
R. Horodecki, P. Horodecki, M. Horodecki, K. Horodecki,
Rev. Mod. Phys. \textbf{81}, 865 (2009)

\bibitem{Pan12}
Jian-Wei Pan, Zeng-Bing Chen, Chao-Yang Lu, H. Weinfurter,
A. Zeilinger, and M. Zukowski
Rev. Mod. Phys. \textbf{84}, 777 (2012)

\bibitem{Schrodinger1935}
E. Schr\"{o}dinger,
Mathematical Proceedings of the Cambridge Philosophical Society, \textbf{31}, 555 (1935)

\bibitem{bell1964}
J. S. Bell, Physics, \textbf{1}, 195 (1964)

\bibitem{aspect1982} A. Aspect, P. Grangier, and G. Roger,
Phys. Rev. Lett. \textbf{49}, 91, (1982)

\bibitem{TrepsFabre} 	
N. Treps and C. Fabre, Laser Phys. \textbf{15}, 187, (2005)

\bibitem{Harold}  H. Ollivier and W. H. Zurek, Phys. Rev. Lett. \textbf{88}, 017901, (2001)

\bibitem{Kavan} K. Modi, A. Brodutch, H. Cable, T. Paterek, Rev. Mod. Physics, \textbf{84} 1655 (2012)

\bibitem{scully04}
A.  Muthukrishnan, G. S. Agarwal and M. O. Scully, Phys. Rev. Lett. \textbf{93}, 093002 (2004)

\bibitem{TP-G-Mayer}
 M. G\"opert-Mayer, Ann. Phys. \textbf{401},  273 (1931)
\bibitem{Mollow}
 B. R. Mollow, Phys.Rev \textbf{175}, 1555 (1968)
\bibitem{Bjorkholm}
J. E. Bjorkholm and P. F. Liao, Phys. Rev. Lett. \textbf{33}, 128 (1974)
\bibitem{Fei}
H.-B. Fei, B. M. Jost, S. Popescu, B. E. A. Saeh and M. C. Teich, Phys. Rev. Lett. \textbf{78}, 1679 (1997)
\bibitem{Lloyd}
S. Lloyd, Science \textbf{321}, 1463 (2008)
\bibitem{Kastella}
K. Kastella and R. S. Conti,
Phys. Rev. A \textbf{83}, 014302 (2011)
\bibitem{Jose}
J. R. Rios Leite, Cid B. De Araujo, Chem. Phys. Lett. {\bf 73}, 71 (1980)


\bibitem{white}
J. C. White, Opt. Lett.  \textbf{6}, 242 (1981)
\bibitem{Sandoghdar}
 Hettich, C. Schmitt, J. Zitzmann, S. K\"{u}hn, I. Gerhardt, V. Sandoghdar,
Science  {\bf 11}, 385 (2002)
\bibitem{bagnato} E. Pedrozo-Pe\~nafiel, R. R. Paiva, F. J. Vivanco, V. S. Bagnato, and K. M. Farias,
Phys. Rev. Lett. \textbf{108}, 253004 (2012)

\bibitem{agarwal92}
M. S. Kim and G. S. Agarwal, Phys. Rev. A  \textbf{57}, 3059, (1998)

\bibitem{Salikhov}
M. Salikhov,  Appl. Magn. Reson. {\bf 25}, 261 (2003)

\bibitem{Mukamel}
 M. Richter, S. Mukamel, Phys. Rev. A. \textbf{83}, 063805 (2011)
 
 \bibitem{Dayan}B. Dayan,
Phys. Rev. A \textbf{76}, 043813 (2007)

\bibitem{Andrews83}
D. L. Andrews and M. J. Harlow, Chem. Phys. Lett. {\bf 78}, 1088 (1983)

\bibitem{Duan} LuMing Duan, G. Giedke, P. Zoller, I. Cirac, Phys. Rev. Letters \textbf{84} 2722 (2000)

\bibitem{FabreMMQO} G. Grynberg, A. Aspect, C. Fabre, \textit{Introduction to Quantum Optics}, (Cambridge University Press, Cambridge, 2010).

\bibitem{ScullyQObook} M. O. Scully and M. S. Zubairy,  \textit{Quantum Optics}, (Cambridge University Press, Cambridge, 1997).

\bibitem{Howell}
I. Khan, J. Howell, Phys. Rev. A  \textbf{73}, 031801(R) (2006)

 \bibitem{wang2006} Kaige Wang, J. Phys. B: At. Mol. Opt. Phys. \textbf{93}, R293 (2006).
 
 \bibitem{handbook} I. S. Gradshteyn and L. M. Ryzhik, page 488, Tables of Integrals, Series, and Products (Academic, New York, 1980).
 
 \bibitem{Barak}
B. Dayan, Y. Bromberg, I. Afek, and Y. Silberberg, Phys. Rev. A  \textbf{75}, 043804 (2007)


\bibitem{Loudon} R. Loudon, \textit{The Quantum theory of light} (Oxford Science Publishing, Oxford, 2000).



\end{thebibliography}
\end{document}